\documentstyle[12pt,epsfig,multicol]{article}
\begin{document}
\def\up#1{\raise 1ex\hbox{\small#1}}
\def\do#1{\lower 1ex\hbox{\small#1}}

\begin{center}
{\LARGE \bf Three sided complex adaptative systems}

\vskip 0.6cm
{\large R.D'hulst \footnote{corresponding author: Rene.DHulst@brunel.ac.uk} and G.J. Rodgers}
\vskip 0.4cm

{\it Department of Mathematics and Statistics, Brunel University,\\
Uxbridge, Middlesex, UB8 3PH, UK.}

\end{center}
\setlength{\unitlength}{1cm}

{\noindent \large Abstract}
\vskip 0.4cm

We introduce two three sided adaptative systems as toy models to mimic the exchange of commodities between buyers and sellers. These models are simple extensions of the minority game, exhibiting similar behaviour as well as some new features. The main difference between our two models is that in the first the three sides are equivalent while in the second, one choice appears as a compromise between the two other sides. Both models are investigated numerically and compared with the original minority game. 


\begin{multicols}{2}
\section{Introduction}
\label{sec:introduction}

\indent\indent In this paper, we consider extensions of the bar-attendance problem introduced by Arthur \cite{arthur94} and simplified into a minority game by Challet and Zhang \cite{challet9798}. In its simplest form, the minority game mimics the internal dynamic of the exchange of one commodity. Agents are allowed to buy or sell this commodity at each time step. No attempt is made to model any external factors that influence the market. Here, we introduce symmetric and asymmetric three sided games as extensions of the minority game. 

In the symmetric three sided model, the agents have to choose between three identical sides at each time step.  These three sides are trading with each other, agents on one side buying from the second side to sell to the third. This model mimics the cyclic trading of goods. If we group any two sides together and consider the trading between this imaginary group and the third side, the model reduces to a kind of minority game with an uneven distribution of the agents. Hence, the connection between this model and the minority game is very strong.

In the asymmetric three sided model, the agents can buy or sell a commodity at each time step, but they can also be inactive, that is, they are allowed to miss a turn. In contrast to the symmetric model, the three choices are not equivalent, as being inactive appears as a compromise between buying and selling. This model can be thought of as an open minority game in the sense that the agents buying and selling are playing a minority game with a variable number of agents at each turn.

In Sec. \ref{sec:themodels}, the minority game is briefly recalled and the two new three sided models are described in detail. In Sec. \ref{sec:symmetric}, the symmetric three sided model is numerically investigated, while in Sec. \ref{sec:asymmetric}, the asymmetric three sided model is investigated. Sec. \ref{sec:conclusions} presents a comparison between the minority game and the two three sided models, as well as our conclusions.

\section{The models}
\label{sec:themodels}

\indent\indent In the minority game, an odd number $N$ of agents have to choose between two sides, $1$ or $2$, at each time step. An agent wins if he chooses the minority side. The record of which side was the winning side for the last $m$ time steps constitutes the history of the system. The agents analyze the history of the system in order to make their next decision.

In the symmetric three sided model, a number of agents $N$ have to choose between three sides, $1$, $2$ or $3$, at each time step. $N$ is not a multiple of 3. The agents on side 1 buy from side 2 to sell to side 3, the agents on side 2 buy from side 3 to sell to side 1 and the agents on side 3 buy from side 1 to sell to side 2. This cyclic trading pattern is shown in Fig. 1. It is assumed that the profit or loss at a side is reflected in the difference between the number of agents they are selling to and the number of agents they are buying from. For instance, $N_3 -N_2$ is a measure of the profit of side 1. Agents choosing the side with the highest profit win and are rewarded with a point. Agents choosing the side with the lowest profit lose and consequently lose a point. Agents choosing the side with the intermediate profit neither lose nor gain a point. Agents strive to maximize their total number of points. 
\end{multicols}
\begin{figure}[h]
  \centering\begin{picture}(5,5)(-2,1)
\put(-0.2,4.9){$N_1$}
\put(0,5){\circle{2}}
\put(-1.932,1.9){$N_2$}
\put(-1.732,2){\circle{2}}
\put(1.532,1.9){$N_3$}
\put(1.732,2){\circle{2}}
\put(0.732,2){\vector(-1,0){1.364}}
\put(-1.232,2.866){\vector(1,2){0.682}}
\put(0.5,4.134){\vector(1,-2){0.682}}
\put(-3.5,5){buying from 2}
\put(-4,3){selling to 1}
\put(-3,0.8){buying from 3}
\put(1,0.8){selling to 2}
\put(2,3){buying from 1}
\put(1,5){selling to 3}
\end{picture}

  \caption{Schematic representation of the symmetric three sided model. The arrows indicate the direction of the exchange. $N_i$ is the number of agents choosing side $i$.
 \label{fig1}}
\end{figure}
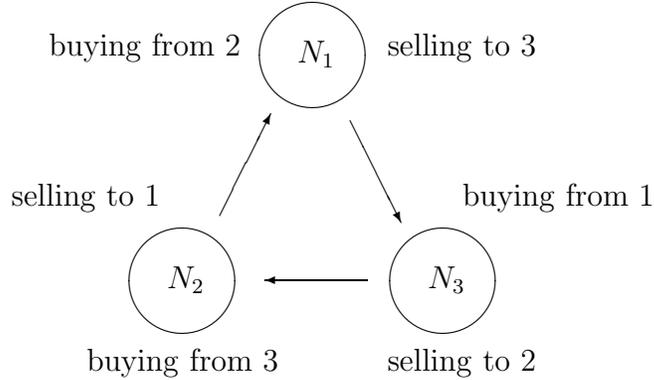

\begin{multicols}{2}
In the asymmetric three sided model, a number of agents $N$ also have to choose between three sides, $1$, $2$ or $3$, at each time step. $1$ corresponds to selling, $2$ to doing nothing and $3$ to buying. The agents buying or selling are said to be active, while the agents doing nothing are said to be inactive. The agents choosing the smaller group among buyers and sellers win and are rewarded with a point. The agents choosing the larger group among buyers and sellers lose and they lose a point. The points of the inactive agents don't change. If there is the same number of buyers and sellers, the points of all the agents remain unchanged. However, the inactive agents are recorded as winners in the history of the system, on the grounds that they achieved the same result as the buyers and sellers, but without taking any risk. Again, agents strive to maximize their total number of points. 
\end{multicols}
\begin{figure}
 \begin{minipage}[b]{.46\linewidth}
\begin{center}
\begin{tabular}{|c|c|c|}
\hline
History & $\sigma$  & $\sigma'$\\
\hline
(1,1) & 3 & 2 \\
(1,2) & 1 & 3 \\ 
(1,3) & 2 & 2 \\ 
(2,1) & 2 & 1 \\
(2,2) & 3 & 3 \\
(2,3) & 1 & 1 \\
(3,1) & 3 & 1 \\
(3,2) & 2 & 2 \\
(3,3) & 3 & 1 \\
\hline
\end{tabular}
\end{center}
\vfill

 \end{minipage} 
 \hfill
 \begin{minipage}[b]{.46\linewidth}
\noindent Table 1: The first column lists all the possible histories of the system for the last 2 time steps $(m=2)$. A strategy is a set of decisions for all the different possible histories. Two example strategies $\sigma$ and $\sigma'$ are shown in the second and third columns.
 \end{minipage}
\end{figure}

\begin{multicols}{2}
In each model, the record of which side was the winning side for the last $m$ time steps constitutes the history of the system. For a given $m$, there are $3^m$ different histories. The 9 different histories for $m=2$ are listed in the first column of table 1. Every agent makes a decision for the next time step according to the history of the system. To be able to play, an agent must have a strategy that allows him to make a decision for any of the $3^m$ different histories. The second and third columns of table 1 list two possible sets of decisions, $\sigma$  and $\sigma'$, that we will call strategies.

Each agent has at his disposal a fixed set of $s$ strategies chosen at random, multiple choices of the same strategy being allowed. At any one moment in time, the agent only uses one of these strategies to make a decision. To allow an agent to decide which strategy to use, every strategy is awarded points, which are called virtual points. The virtual points of a strategy are the points the agent thinks he could have earned had he played with this strategy. Hence, the virtual points are rewarded using the same scheme as the points given to the agents, the prediction of a strategy being compared to the actual decisions. A strategy predicting the winning side is awarded a virtual point, a strategy predicting the losing side loses a virtual point and a strategy predicting the third side does not gain or lose any points. In the asymmetric model, in the case of an equal number of buyers and sellers, the virtual points of all strategies remain unchanged. An agent always plays with the strategy with the highest number of virtual points. When more than one strategy has the highest number of virtual points, one of them is chosen at random. 

If we compare two strategies $\sigma$ and $\sigma'$ component by component, we see that for some histories they can make the same prediction and for others they can make different predictions. In the example in table 1, the decisions differ when the history is (1,1), (1,2), (2,1), (3,1) and (3,3). To consider this feature, we have to distinguish between the symmetric model and the asymmetric one. For the former, the three sides are equivalent and only the number of differences between the strategies can give a measure of the difference between two strategies in the strategy space. For the latter, there is a qualitative difference between the three sides. This qualitative difference should appear in the definition of the difference between strategies. 

Consider first the symmetric three sided model. As the three sides are equivalent, a geometrical representation should put them at the same distance from one another. A convenient measure of the differences between two strategies $\sigma$ and $\sigma'$ is 

\begin{equation}
d_s = {1\over 3^m} \sum_{i=1}^{3^m} \delta (\sigma_i - \sigma'_i )
\label{eq:symmetric distance definition}
\end{equation}
where $\delta (0) = 1$, and $\delta (x) = 0$ otherwise. $d_s$ is defined as the distance between strategies in the symmetric model. This definition takes into account the geometrical structure of the strategy space, including the equivalence between the three sides. In the example of table 1, $d_s =5/9$. By definition, the symmetric distance is a number ranging from 0 to 1.

As Eq. (\ref{eq:symmetric distance definition}) shows, the symmetric distance $d_s$ is defined as a sum of $3^m$ terms, which we label $d_s^{(i)}$'s. Each of these terms is equal to 0 with probability 1/3 or equal to 1 with probability 2/3. The average distance between two strategies is $\overline{d}_s = 2/3$, while the variance of the symmetric distance distribution is $\sigma^2_s  = 2/3^{m+2}$. The symmetric distance between two strategies corresponds to the probability that these two strategies will give different predictions, assuming that all the histories are equally likely to occur. The symmetric distance corresponds to the distance defined in the minority game \cite{dhulst99}. Two strategies at $d_s = 0$ are correlated, two strategies at $d_s  =2/3$ are uncorrelated and two strategies at $d_s = 1$ are anticorrelated. 

In the asymmetric three sided minority game, selling is just the opposite decision to buying while doing nothing is a compromise. Consequently, the normalized asymmetric distance between strategies $\sigma$ and $\sigma'$, 

\begin{equation}
d_a = {1\over 3^m} \sum_{i=1}^{3^m} {| \sigma_i - \sigma'_i |\over 2}
\label{eq:asymmetric distance definition}
\end{equation}
is a measure of the difference between the two strategies. $d_a$ is defined as the distance between strategies in the asymmetric model. This definition takes into account the fact that buying is more different from selling than it is from being inactive in an arbitrary way. In the example of table 1, $d_a =4/9$. By definition, the asymmetric distance is a number ranging from 0 to 1.

As shown by Eq. (\ref{eq:asymmetric distance definition}), the asymmetric distance $d_a$ is defined as a sum of $3^m$ terms we label $d_a^{(i)}$'s. When the component of a strategy is equal to 2, this component can never give a $d_a^{(i)}$ equal to 1. In other words, the inactive side has no side at distance 1 from itself. Considering all the possibilities, the probability to find a $d_a^{(i)}$ of 0 is 1/3, of 0.5 is 4/9 and of 1 is 2/9. The average asymmetric distance between strategies is $\overline{d}_a = 4/9$, while the variance of the asymmetric distance distribution is equal to $\sigma^2_a = 11/3^{m+4}$. The interpretation of this asymmetric distance is ambiguous. In fact, the opposite to selling is buying, but the opposite to being inactive is being inactive. Hence, $d_a$ is not a measure of the probability that two strategies would give opposite decisions.  Two correlated strategies are at $d_s = 0$ from each other, two uncorrelated strategies are at $d_a = 1/2$ from each other, but two anticorrelated strategies can be at $d_a = 0$ or $d_a = 1$ from each other. 

\section{Numerical results for the symmetric model}
\label{sec:symmetric}

\indent\indent In this section, we report on numerical investigations of the properties of the symmetric three sided model, interpreting the results using the symmetric distance.

Fig. 2 presents a typical result for the time evolution of the attendance at one side. The simulation is for $N=101$ agents with $s=2$ strategies each and a memory of $m=3$. The result for the attendance at one side is very similar to the results of the minority game, the mean attendance being shifted to $N/3$ instead of $N/2$. Given an agent choosing one side, the average distance between the strategy used by this agent and the strategies used by the other agents is $\overline{d} = 2/3$. That is, around $2/3$ of the agents should choose one of the two other sides. Hence, the average attendance at one side is $N/3$.
\end{multicols}
\begin{figure}[h] 
   \centerline{\epsfig{file=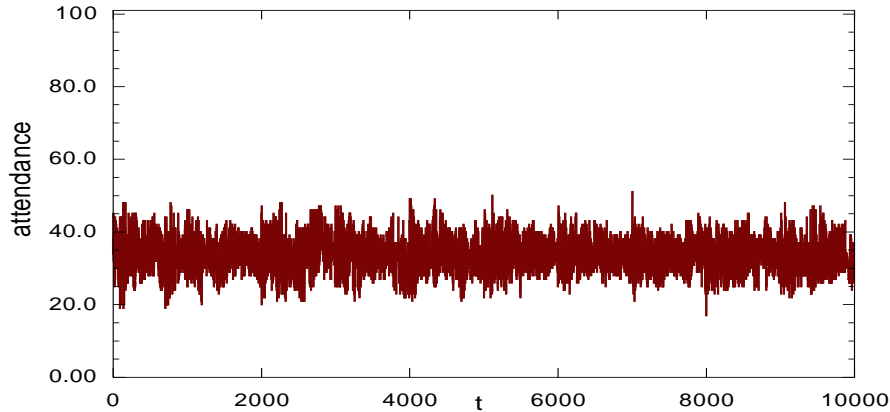, 
   width=15cm,height=8cm}}
   \caption{Numerical simulation of the attendance at one side for the symmetric three sided model. The choice for the parameters is $N=101$, $s=2$ and $m=3$.}
   \label{fig2} 
\end{figure} 

\begin{multicols}{2}
The variance of the attendance at one side as a function of the size of the memory $m$ is presented at Fig. 3 for $N=101$ agents with $s=2$ strategies. The result is again very similar to the minority game, with a very high variance for $m<3$, a minimum at $m=3$ and the variance going to $2N/9$, the random value, as $m$ goes to infinity. Curves of the same shape are obtained for the variance of the number of winners or the variance of the number of losers. Also, the maximum profit or the number of agents on the more crowded side exhibit the same behaviour. Each of these curves has a minimum for $m=3$. For $m<3$, the number of strategies used at each time step is a representative sample of the space of the strategies. Consequently, the variance of the attendance is directly related to the variance of the distance distribution, $\sigma^2_s  = 2/3^{m+2}$. In fact, the variance of the attendance scales like $1/ \sigma^2_s$. On the contrary, for $m>3$, the space of the strategies is very large, so that most of the strategies used are uncorrelated. As a result, the kinetics of the system are the same as the kinetics of a random walk. Between these two behaviours, for $m$ around 3, the agents organize themselves better, a crowd-anticrowd effect being obtained \cite{johnson98-1}.
\end{multicols}
\begin{figure}[h] 
   \centerline{\epsfig{file=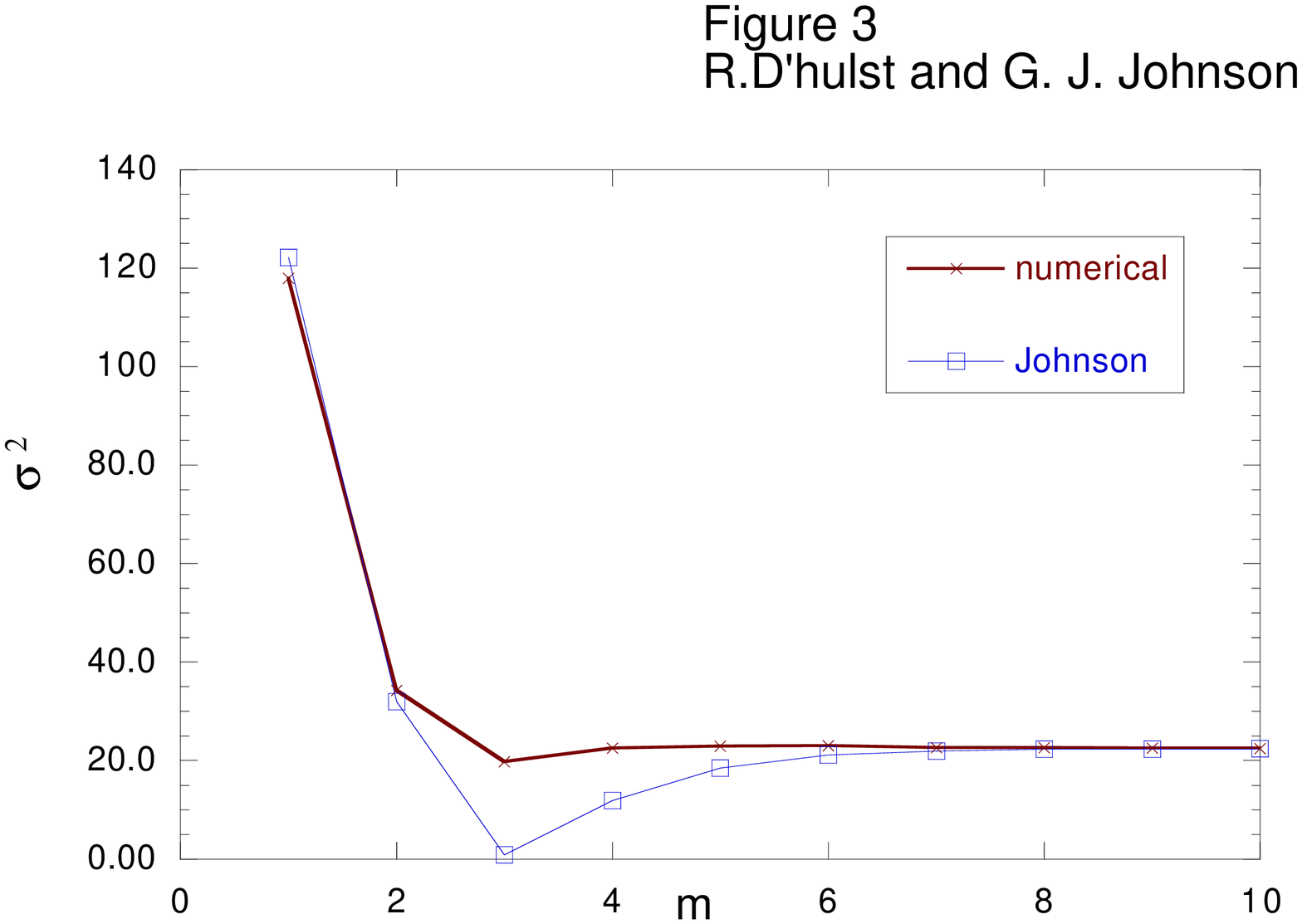, 
   width=15cm,height=8cm}}
   \caption{Variance of the number of agents at one side as a function of the size of the memory $m$ ($\times$) for $N=101$ and $s=2$ (symmetric model). The analytical results using the calculation of Johnson {\it et al.} [4] is also shown ($\Box$).}
   \label{fig3} 
\end{figure} 

\begin{multicols}{2}
Even for small values of $m$, the space of strategies is very large, of size $3^{3^m}$. But as in the minority game, not all the strategies are uncorrelated. If we suppose that $1/ \sigma^2_s$ gives an estimate of the number of uncorrelated strategies, the method of Johnson {\it et al.} \cite{johnson98-1} can be used to find an analytical expression for the variance of the attendance at one side. We followed the original calculation in \cite{johnson98-1}, with a size $a = 3^{m+2}/2$ for the space of strategies and a variance of 2/9 for an independent agent. The analytical result obtained by this method is compared in Fig. 3 to the result of the numerical simulations. The curves agree qualitatively. 
\end{multicols}
\begin{figure}
\begin{tabular}{cc}
\begin{picture}(6,2)(0,0)
   \epsfig{figure=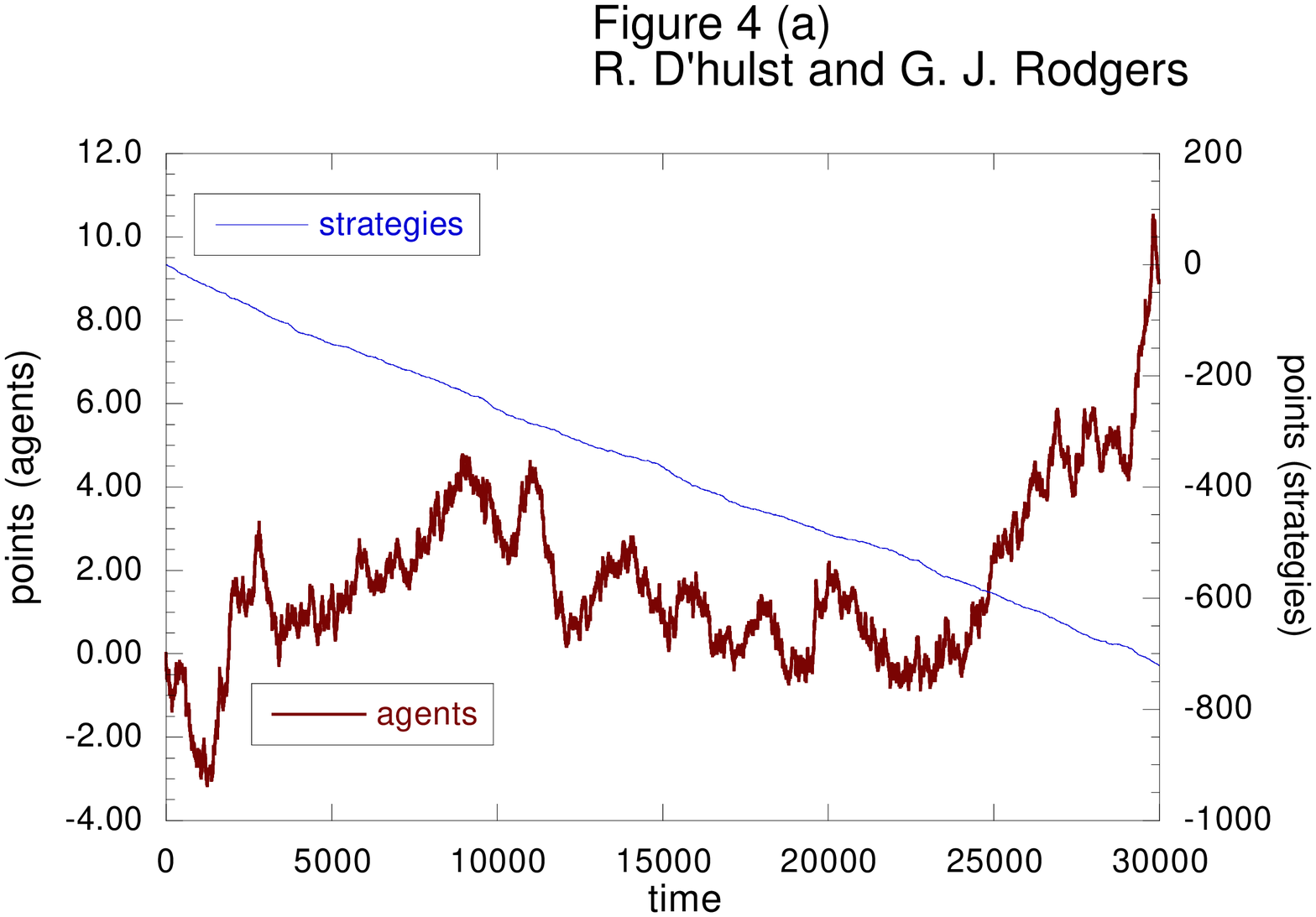,width=0.5\linewidth}
\end{picture}
   &
\begin{picture}(8,2)(0,0)
   \epsfig{figure=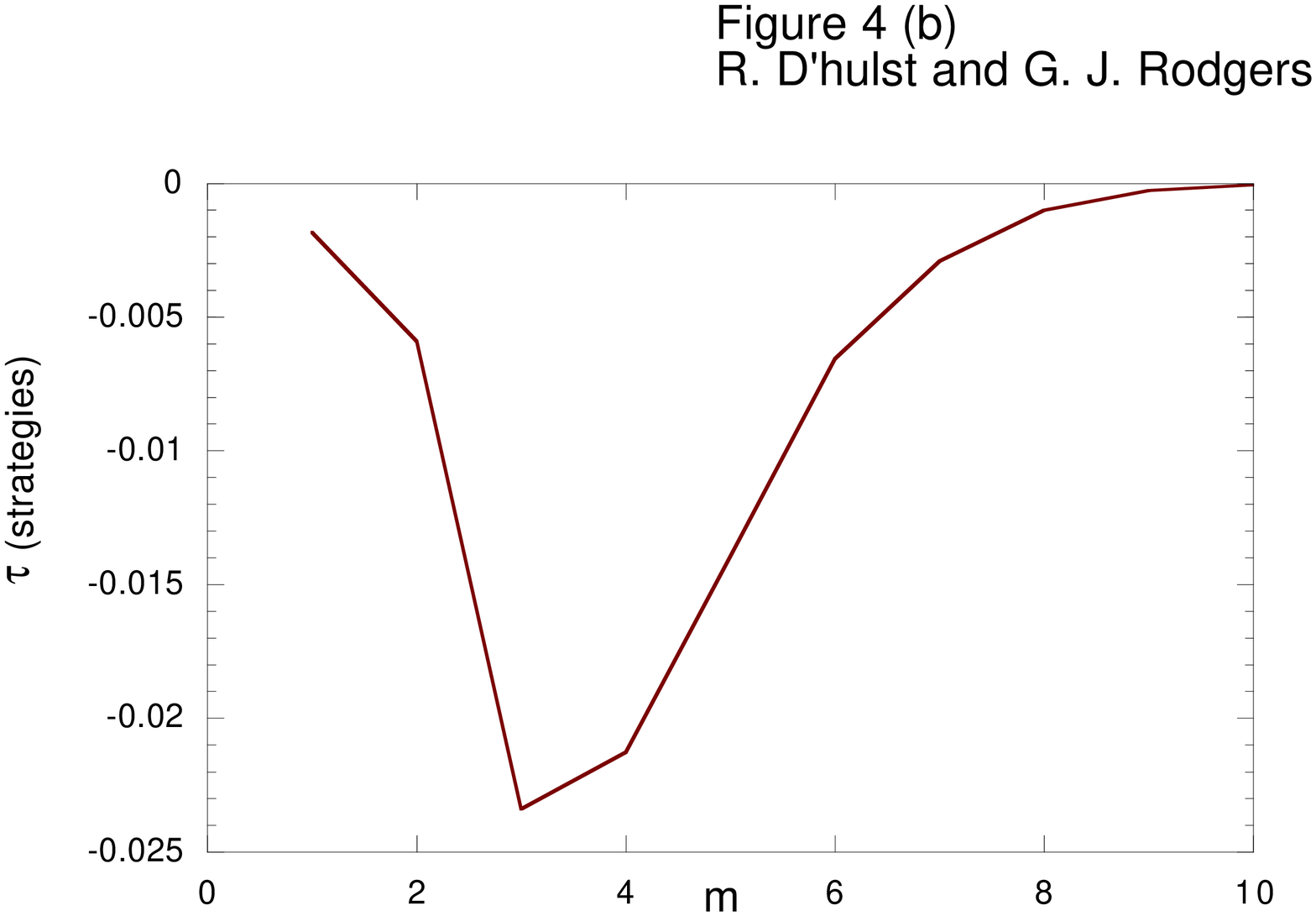,width=0.5\linewidth}
\end{picture}
\end{tabular}
\caption{(a) Average points earned by the agents and their strategies as a function of the time for $N=101$, $s=2$ and $m=3$. The ordinate on the left refers to the points of the agents (bold line) while the ordinate on the right refers to the points of their strategies (simple line). (b) Profit rate of the strategies as a function of $m$ for $N=101$ and $s=2$.}
\label{fig4}
\end{figure}

\begin{multicols}{2}
Fig. 4 (a) presents a typical result for the average number of points given to the agents and their strategies. The parameters of the simulation are $N=101$, $s=2$ and $m=3$. Note that there are two different ordinate scales. As Fig. 4 (a) shows, the virtual points are steadily decreasing with time. In contrast, the points given to the agents display a more complex behaviour. The points given to the agents increase very slowly for $m<3$ and then oscillate around 0 for $m>3$. There seem to be no special behaviour for $m=3$.  The time evolution of the virtual points can be approximated by a linear relation with a profit rate $\tau$. We define $\tau$ as the average number of points earned by a strategy at each time step. Fig. 4 (b) presents $\tau$ as a function of the memory $m$ for $N=101$ and $s=2$. For $m<3$, the strategies are slowly losing points, the worst results being obtained for $m=3$. For $m>3$, the virtual points oscillate around 0. Hence, the agents seem to be able to choose their strategy efficiently, in the sense that the strategies they choose win more often than the average strategy. This behaviour is to be contrasted with the minority game where the agents are not able to choose a strategy efficiently.

As a summary, the symmetric model is a direct extension of the minority game to three sides. The results found are very similar, with a glassy phase transition \cite{savit97} when the size of the memory of the agents is increased. We numerically identified a critical value $m_c$ for the size of the memory. For $m<m_c$, the space of strategies is crowded and its geometrical structure is apparent in the results. As this structure is encoded in the distance definition, the system is driven by its distance distribution . For $m>m_c$, the number of strategies used is not relevant as most of the strategies used are uncorrelated. The kinetics of the system reduce to agents choosing one of the three sides at random. Hence, there is a transition from a system driven by its distance distribution to a random system. 

\section{Numerical results for the asymmetric model}
\label{sec:asymmetric}

\indent\indent We investigated numerically the different properties of the asymmetric three sided game. In the figures, 1 denotes buying, 2, doing nothing and 3, selling.

The attendance of the three different sides as a function of the size of the memory $m$ is plotted at Fig. 5 for $N=101$ agents, playing with $s=2$ strategies each. The number of agents in the winning side is also presented. For small $m$ values, most of the agents are buying or selling (the two superimposed upper curves). Just a few of them are doing nothing (the lower curve for small $m$ values). As the size of the memory is increased, the system corresponds more and more to the agents guessing at random between the three possibilities. Also, for small values of $m$, the number of winners is significantly more than 1/3, the random guess value. Fig. 5 is interesting because the difference between the three sides is clearly apparent.  
\end{multicols}
\begin{figure}[h] 
   \centerline{\epsfig{file=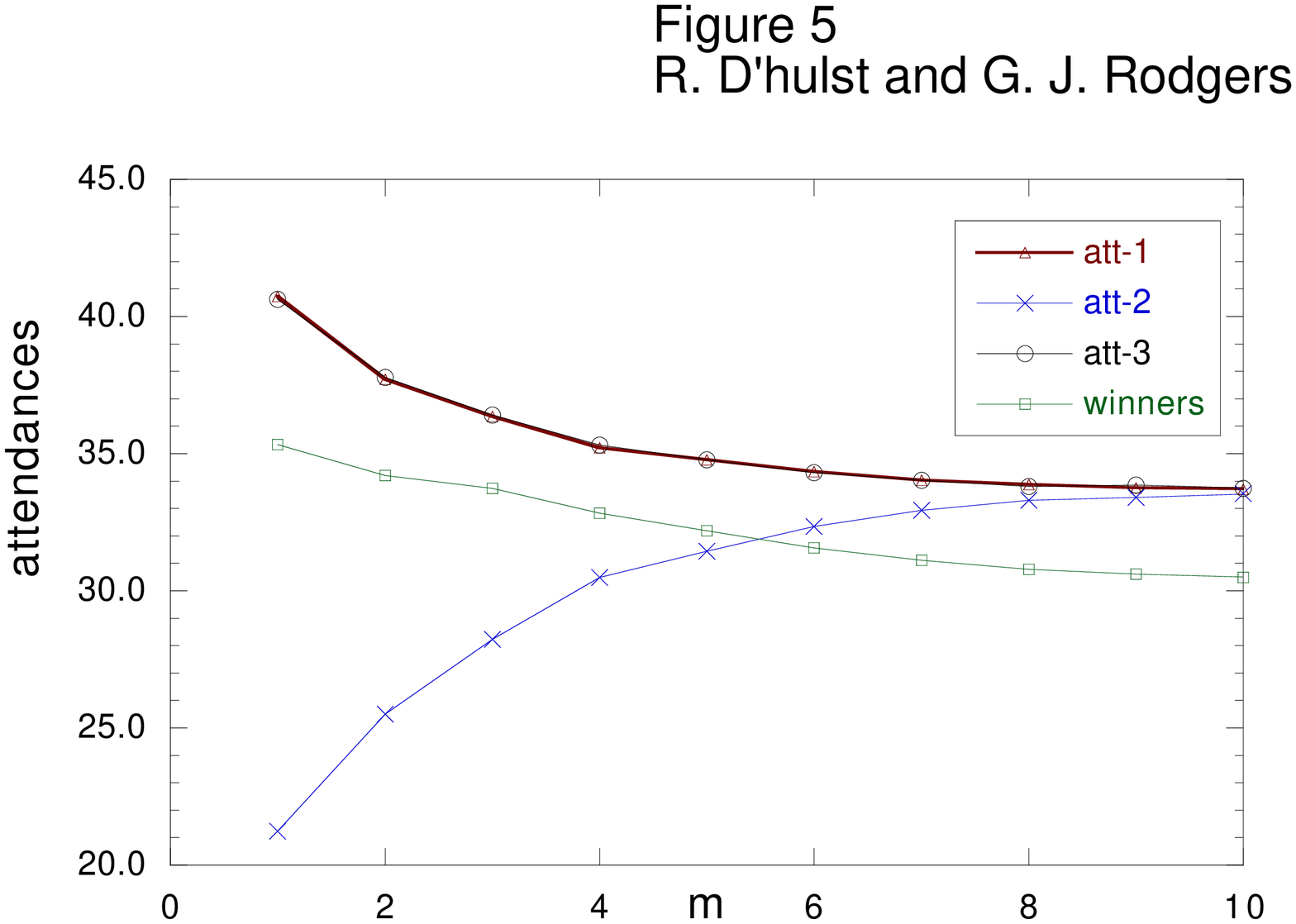, 
   width=15cm,height=8cm}}
   \caption{Numerical result for the average attendances at the three sides as functions of $m$ for $N=101$ and $s=2$. The two upper curves ($\triangle$ and $\bigcirc$) are for the selling and buying options. The lower curve ($\times$) is for the inactive agents. The average number of winners is also presented ($\Box$) as a function of $m$ for the same choice of parameters.}
   \label{fig5} 
\end{figure} 

\begin{multicols}{2}
In Fig. 6, the variance of the attendances at the three sides and the variance of the number of winners are presented as functions of $m$ for $N=101$ and $s=2$. For $m<6$, the variance of the number of inactive agents is significantly higher than $2N/9$, the value for agents guessing at random. The variances of the number of buyers and sellers has a minimum at $m=2$.  The variances of the three sides increase to $2N/9$ as $m$ increases. Hence, there seems to be an organization of the agents for $m$ around 2. The variance of the number of winners has a shape very similar to the one found in the minority game. For small value of $m$, the variance diverges like a power law of $m$; at $m\simeq 7$, it seems to reach a minimum and for higher values of $m$, it goes asymptotically to a value near $N/9$. However, the existence of a minimum at $m=7$ could not be confirmed unequivocally by the numerical simulations. 
\end{multicols}
\begin{figure}[h] 
   \centerline{\epsfig{file=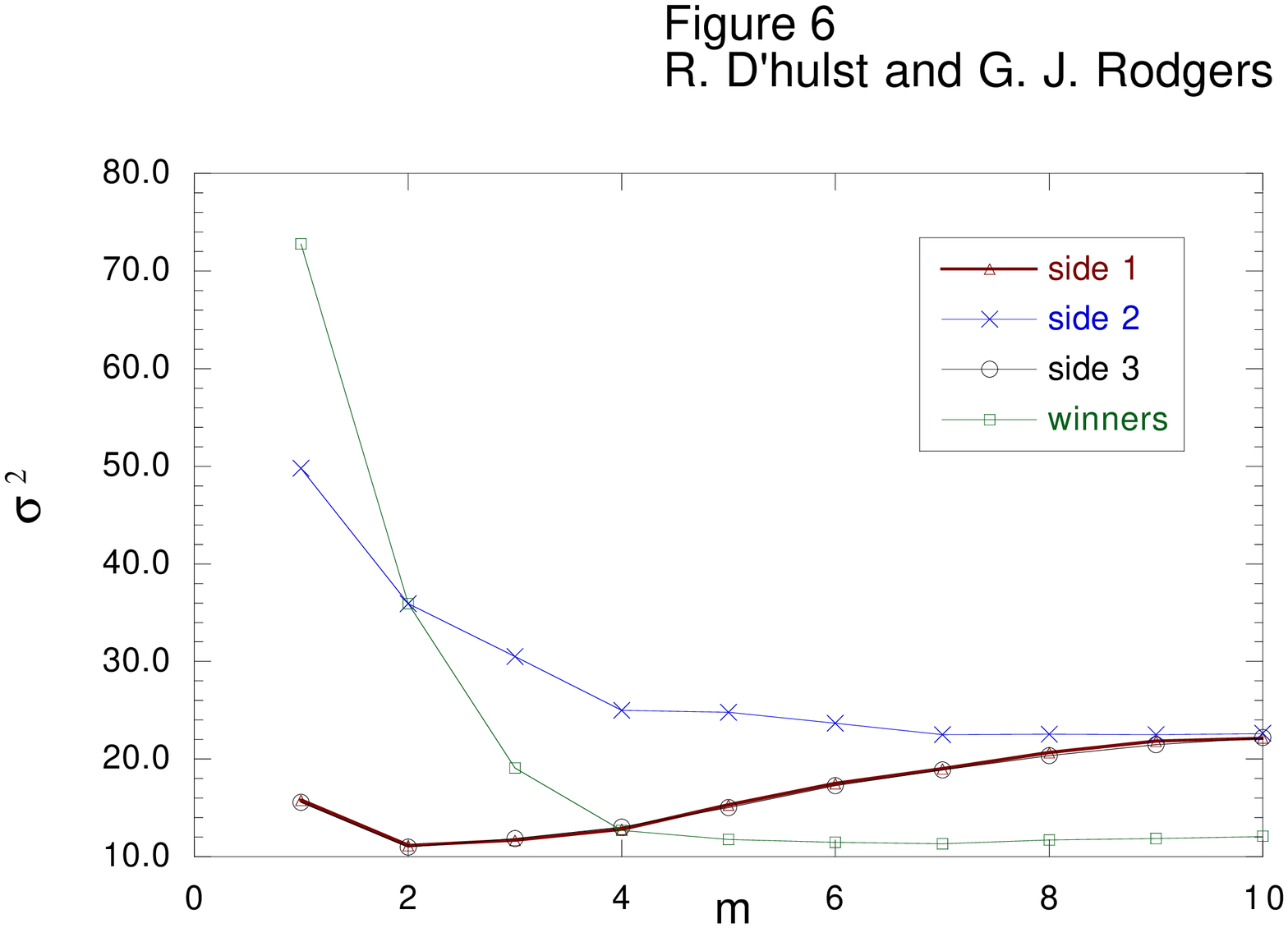, 
   width=15cm,height=8cm}}
   \caption{Variances of attendances at the three sides as functions of $m$ for $N=101$ and $s=2$. $\triangle$ and $\bigcirc$, selling and buying options, $\times$ inactive agents. $\Box$ is the variance of the number of winners as a function of $m$ for the same choice of parameters.}
   \label{fig6} 
\end{figure} 

\begin{multicols}{2}
Fig. 5 and 6 show that for small $m$ values, the behaviour of the system is directly related to the properties of the distance distribution. The proportion of people buying or selling is of the same order as the average distance, 4/9, while the variance of the number of winners scales as $1 /\sigma^2_a$, the inverse of the variance of the asymmetric distance distribution. These properties were also present in the minority game. In this asymmetric model, the variance of the attendance at one side does not represent the wasted number of points. The wasted number of points is defined to be the difference between the maximum points that can be earned by the system at each time step and the average points actually earned by the system at each time step. This is why we also have to consider the properties of the number of winners in addition to the properties of the attendances.

For higher values of $m$, the strategy space is so large that most of the used strategies are uncorrelated. The system is similar to a system with agents choosing at random from the three sides. In the minority game, the relative attendance predicted by the distance distribution is the same as the one predicted by random guesses, that is 1/2. In the present three sided minority game, these two ratios are 4/9 and 1/3 respectively. Hence, the transition from a system driven by the distance distribution to a system of agent guessing at random is seen directly in the attendance of the different sides.

Fig. 7 presents the average success rate of one side, that is, the probability that at any one moment in time, one side will win. As expected, the sides corresponding to buying and selling are symmetric and more likely to win than the inactive side. In fact, there are $(N+1)(N+2)/2$ different configurations for the attendances of the 3 sides. Among these, only $(N+2)/2$ make the inactive agents winners if $N$ is even, $(N+1)/2$ if $N$ is odd. Hence, if all the situations were equally likely to occur, the inactive agents would win at most about every $N+1$ time steps. This is the order of the asymptotical value for the success rate of this side. For low values of $m$, the success rate of the inactive side is higher than the asymptotical value, implying that the agents playing are organizing themselves rather well. The transition between organized and non-organized agents is for $m=2$ in Fig. 7.
\end{multicols}
\begin{figure}[h] 
   \centerline{\epsfig{file=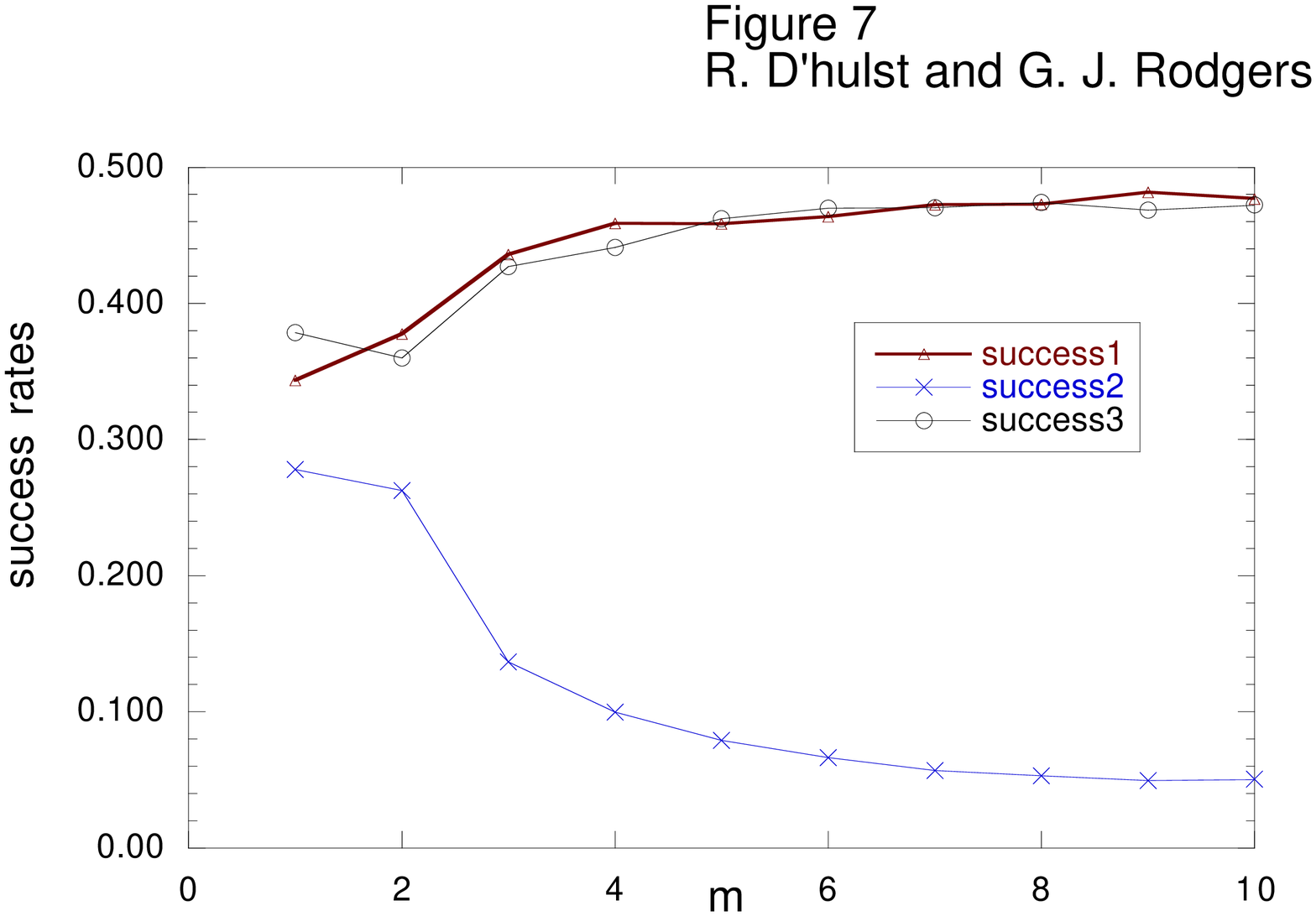, 
   width=15cm,height=8cm}}
   \caption{Success rates of the three different sides as functions of $m$ for $N=101$ and $s=2$. The two upper curves ($\triangle$ and $\bigcirc$) are for selling and buying while the lower curve ($\times$) is for the inactive agents.}
   \label{fig7} 
\end{figure} 

\begin{multicols}{2}
Fig. 8 confirms the organization of the agents. The profit rates of the agents and their strategies are shown as functions of $m$. We define a profit rate as the average number of points earned at each time step. For values of $m$ less than $m=5$, the agents are able to choose strategies which are more successful than the average ones. On the contrary, for $m>5$, they are doing worse than guessing at random. The curve of the profit rate of the strategies suggests that the transition takes place for $m=2$.
\end{multicols}
\begin{figure}[h] 
   \centerline{\epsfig{file=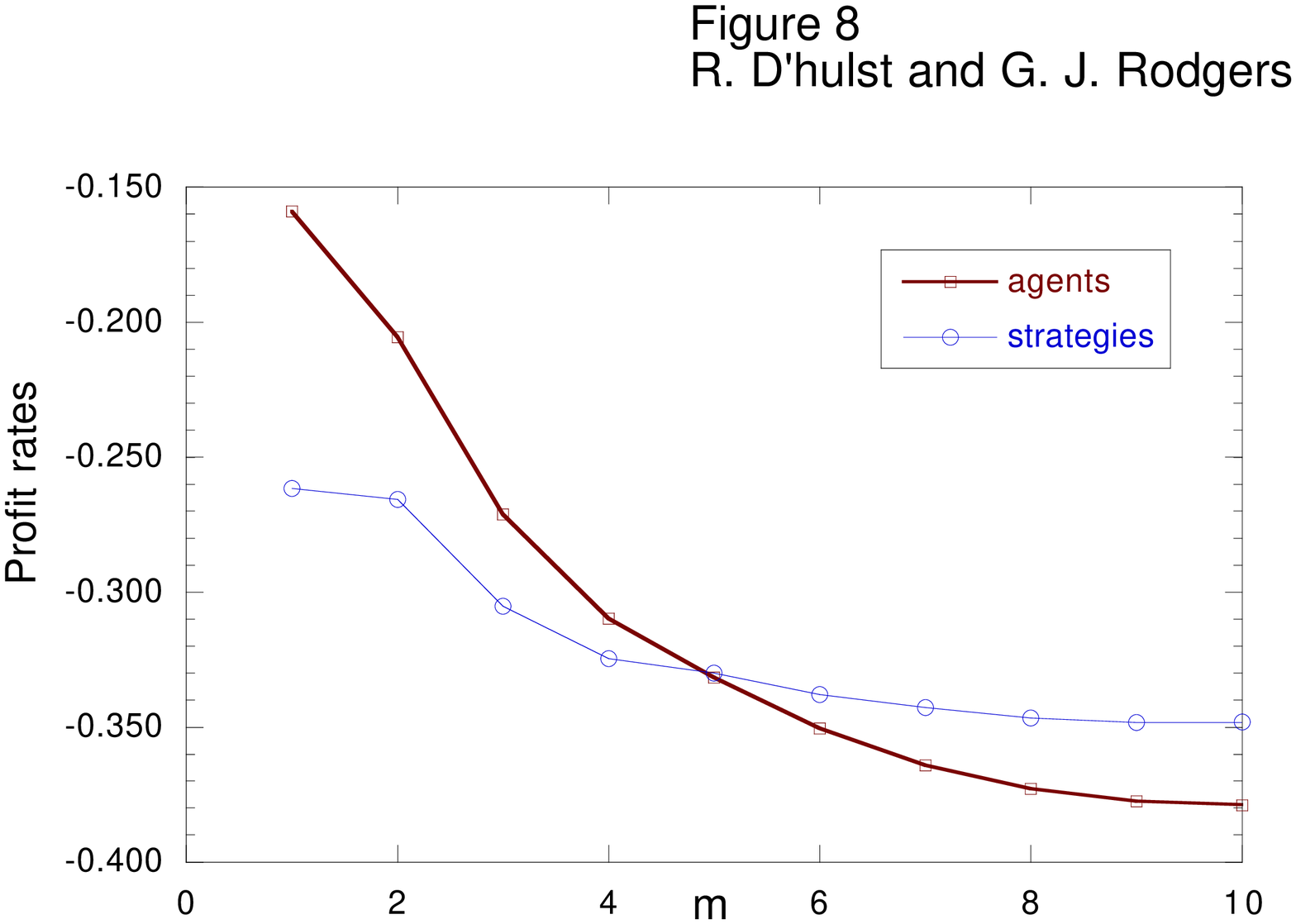, 
   width=15cm,height=8cm}}
   \caption{Average profit rates of the agents and their strategies as functions of $m$ for $N=101$ and $s=2$. $\Box$ is for the profit rate of the agents while $\bigcirc$ is for the profit rate of their strategies.}
   \label{fig8} 
\end{figure} 

\begin{multicols}{2} 
As a summary, in the asymmetric three sided minority game, agents playing with a small memory win more points on average than agents playing with a bigger memory in a pure population, that is, a population of all the agents with the same memory size $m$. As in the minority game and the symmetric three sided model, a glassy phase transition \cite{savit97} is found at a particular value of the memory $m_c$. For $m<m_c$, the geometrical properties of the space of strategies is apparent, especially the asymmetry between the three sides. Most of the agents are playing and the system is driven by the distance distribution. In contrast to the minority game, this property is seen directly in the number of agents on each side. For $m>m_c$, the strategies used are uncorrelated and the system is similar to a system of agents guessing at random. Considering the adaptation of the agents, they are unable to realize that the wiser choice is to decide to be inactive. In fact, more than half of the active agents will lose. The agents are fooled because they base their confidence in virtual points, not on their profit. Hence, the agents are always tempted to play even if they are unlikely to win.

\section{Conclusions}
\label{sec:conclusions}

\indent\indent We introduced two three sided models as extensions of the minority game. In the symmetric three sided model, agents are given three equivalent choices while in the asymmetric three sided model, agents have the opportunity to miss a turn and not play. We have investigated these two new models numerically and compared the results with the original minority game.

In both models, we defined a distance between the strategies of the agents. These distances incorporate in their definitions the geometrical structure of the space of strategies. In the symmetric model, the geometrical structure of the space of strategies is very similar to the one in the minority game. The distance gives a measure of the correlation between two strategies. Conversely, the distance in the asymmetric model has no obvious interpretation. 

A transition between a system driven by the distance distribution and a system of agents guessing at random was identified numerically in both models. However, in contrast to the minority game, the agents make their highest profit for $m$ small and not at the transition value of $m$. In the distance driven phase, the agents organize themselves, as in the minority game. In contrast to the minority game, however, the average profit rate of the agents is higher than the average profit rate of the strategies, indicating that the agents are choosing their strategies efficiently. In the symmetric model, the transition is apparent in the variance of the number of agents choosing one side while in the asymmetric model the transition is seen in the number of agents itself. This latter property of the asymmetric model is a direct consequence of the geometrical structure of the space of strategies.

In the future, we intend to investigate both models analytically. The symmetric model, in particular, should be amenable to analytical treatment, perhaps following the methods introduced in \cite{challet99} for the two sided minority game.

\end{multicols}
\end{document}